\documentclass[structabstract]{aa}  
\usepackage{amsmath}
\usepackage{natbib}
\usepackage{graphicx}
\usepackage{txfonts}
\newcommand{\kms}{\hbox{${\rm km}\:{\rm s}^{-1}\;$}}
\newcommand{\kmso}{\hbox{${\rm km}\:{\rm s}^{-1}$}}
\newcommand{\teff}{$T_{\rm eff}\;$}  
\newcommand{\teffo}{$T_{\rm eff}$}

\newcommand{\logg}{$\log{g}\;$}   	
  
\newcommand{\loggo}{$\log{g}$}  	
\newcommand{\mic}{$\xi_{\rm t}$\;}

\newcommand{\cobold}{\ensuremath{\mathrm{CO}^5\mathrm{BOLD}}}

\newcommand{\ftac}{3D -- 1D$_{\rm LHD}$}
\newcommand{\tac}{3D -- $\langle$3D$\rangle$}
\newcommand{\foac}{$\langle$3D$\rangle$ -- 1D$_{\rm LHD}$}
\newcommand{\linfor}{Linfor3D}

\newcommand{\pun}[1]{\,#1}

\bibpunct{(}{)}{;}{a}{}{,} 
\begin{document}
\title{Galactic evolution of oxygen}
\subtitle{OH lines in 3D hydrodynamical model atmospheres}
\titlerunning{Oxygen versus Iron using 3D models}
\authorrunning{Gonz\'alez Hern\'andez et al.}
%
%
\author{J.~I.~Gonz\'alez Hern\'andez\inst{1,2,3} \and P.
Bonifacio\inst{2,3,4} \and H.-G.~Ludwig\inst{2,3,5} \and 
E.~Caffau\inst{3} \and N.~T. Behara\inst{2,3,6} \and B.
Freytag\inst{7,8}}
\offprints{J.~I. Gonz\'alez Hern\'andez}
\institute{
Dpto. de Astrof\'{\i}sica y Ciencias de la Atm\'osfera, Facultad de
Ciencias F\'{\i}sicas, Universidad Complutense de Madrid, E-28040
Madrid, Spain  
\email{jonay@astrax.fis.ucm.es}
\and
Cosmological Impact of the First STars (CIFIST) Marie Curie Excellence
Team [http://cifist.obspm.fr]
\and
GEPI, Observatoire de Paris, CNRS, Universit\'e Paris Diderot; Place
Jules Janssen 92190
Meudon, France \\
\email{[Piercarlo.Bonifacio];[Hans.Ludwig];[Elisabetta.Caffau];[Natalie.Behara]@obspm.fr}
\and
Istituto Nazionale di Astrofisica - Osservatorio Astronomico di
Trieste, Via Tiepolo 11, I-34143 Trieste, Italy
\and
Zentrum f\"ur Astronomie der Universit\"at Heidelberg,
Landessternwarte, K\"onigstuhl 12, D-69117 Heidelberg, Germany 
\and
Institut d'Astronomie et d'Astrophysique, Universit\'e Libre de 
Bruxelles, B-1050 Bruxelles, Belgium
\and
CRAL,UMR 5574: CNRS, Universit\'e de Lyon, \'Ecole Normale Sup\'erieure
de Lyon, 46 all\'ee d'Italie, F-69364 Lyon Cedex 7, France
\email{Bernd.Freytag@ens-lyon.fr}
\and
Istituto Nazionale di Astrofisica - Osservatorio Astronomico di
Capodimonte, Via Moiariello 16, I-80131 Napels, Italy}

\date{Received ... March 2010; accepted ... }
 
\abstract
{Oxygen is the third most common element in the
Universe. The measurement of oxygen lines in metal-poor unevolved
stars, in particular near-UV OH lines, can provide invaluable
information about the properties of the Early Galaxy.} 
{Near-UV OH lines constitute an important tool to derive oxygen
abundances in metal-poor dwarf stars. Therefore, it is important to
correctly model the line formation of OH lines, especially in
metal-poor stars, where 3D hydrodynamical models commonly 
predict cooler temperatures than plane-parallel hydrostatic models 
in the upper photosphere.}
{We have made use of a grid of 52 3D hydrodynamical model atmospheres 
for dwarf stars computed with the code \cobold, extracted from 
the more extended CIFIST grid. The 52 models cover the
effective temperature range 5000--6500\,K, the surface gravity range 
3.5--4.5 and the metallicity range $-3 < [{\rm Fe}/{\rm H}] < 0$.}     
{We determine 3D-LTE abundance corrections in all 52 3D models 
for several OH lines and \ion{Fe}{i} lines of different
excitation potentials. These 3D-LTE corrections are generally negative
and reach values of roughly -1~dex (for the OH 3167 with excitation
potential of approximately 1~eV) for the higher temperatures and
surface gravities.}     
{We apply these 3D-LTE corrections to the individual O abundances derived
from OH lines of a sample the metal-poor dwarf stars reported in 
Israelian et al.(1998, 2001) and Boesgaard et al.(1999) 
by interpolating the stellar parameters of the dwarfs in the grid of 
3D-LTE corrections. 
The new 3D-LTE [O/Fe] ratio still keeps a similar trend as the 1D-LTE, i.e, 
increasing towards lower [Fe/H] values. We applied 1D-NLTE
corrections to 3D \ion{Fe}{i} abundances and still see an
increasing [O/Fe] ratio towards lower metallicites. However,
the Galactic [O/Fe] ratio must be revisited once 3D-NLTE
corrections become available for OH and Fe lines for a grid of 3D
hydrodynamical model atmospheres.
}   


\maketitle

\section{Introduction}
\label{intro}

The metal-poor stars of the Galactic halo provide the
fossil record of the early Galaxy's composition. Dwarf halo stars are
particularly relevant, because their atmospheres are not significantly
altered by internal mixing and provide a unique tracer to constrain
Galactic chemical evolutionary models. Oxygen is a key element in this
scenario, because it is the most abundant element in stars after H and
He. It is produced in the interiors of massive stars by hydrostatic
burning and its content is modified during the explosive
nucleosynthesis in type~II supernovae (SNe) and hypernovae (HNe)
and returned into the interstellar medium. On the other hand, iron is
created by both type~II and type~I SN explosions. However,
type~I SNe progenitors 
have longer lifetimes, which is why the abundance ratio of
[O/\ion{Fe}{i}] can be used to constrain the chemical evolution 
of the Galaxy.

There have been numerous studies of the oxygen abundance in halo
stars. Despite considerable observational and theoretical efforts, the
trend of [O/Fe] ratio\footnote{$[{\rm O}/{\rm Fe}]=\log [N({\rm O})/N({\rm
Fe})]_\star-\log [N({\rm O})/N({\rm Fe})]_\odot$} versus [Fe/H]
is still unclear. 
The analysis of the forbidden line \ion{O}{i} 6300~{\AA}
in giants shows a plateau with [O/Fe$] \sim 0.4-0.5$ in the
metallicity range $-2.5 < [{\rm Fe}/{\rm H}] < -1$ \citep{bar88} and
[O/Fe$] \sim 0.7$ for $-4. < [{\rm Fe}/{\rm H}] < -2.5$ \citep{cay04}. 
A similar behaviour is seen for metal-poor subgiant stars in the range
$-3. < [{\rm Fe}/{\rm H}] < -1.5$ \citep[{[O/Fe]} $\sim 0.4-0.5$,][]{gar06}. 
However, \citet{gar06} already noted that by plotting all
measurements from the [OI] line for dwarfs \citep{nis02}, subgiants
\citep{gar06} and giants \citep{cay04}, the picture changes and an
increasing trend [O/Fe] towards lower metallicities clearly appears.

The near infrared (IR) triplet \ion{O}{i} 7771--5~{\AA} in metal-poor dwarfs and
subgiants \citep{abi89,isr01,gar06} points towards increasing [O/Fe]
values with decreasing [Fe/H], although the O abundances derived from
the near-IR triplet are typically $\sim 0.4-0.7$ higher than those
derived from the forbidden \ion{O}{i} line \citep{ful03}. The OH A-X
electronic lines in the near ultraviolet (UV) provide also higher
[O/Fe] ratios towards lower [Fe/H] values in dwarf stars
\citep{isr98,boe99,isr01,gon08}. However, \citet{gar06} found a
quasi-plateau of [O/Fe] for subgiant stars in the range
$-3. < [{\rm Fe}/{\rm H}] < -1.5$, using Fe abundances determined from
\ion{Fe}{ii} lines. The [O/Fe] ratio shows a negative slope if one 
instead uses the Fe abundances estimated from \ion{Fe}{i} lines, although
with lower [O/Fe] values than those determined for metal-poor dwarfs.
It is advisable to use \ion{Fe}{i} lines instead of \ion{Fe}{ii} lines
to derive the [O/Fe] ratio because of its similar sensitivity to the
surface gravity.

These O-abundance indicators present different complications. The
near-IR \ion{O}{i} triplet is susceptible to non-local
thermodynamical equilibrium (NLTE) effects
\citep[and references therein]{kis01}, with abundance corrections
below 0.2 dex, and is quite sensitive to the adopted \teffo. 
The [OI] is not sensitive to departures from LTE, but it is essentially
undetectable in dwarfs with $[{\rm Fe}/{\rm H}]\lesssim -2$. 
The near-UV OH lines 
are strongly sensitive to the temperature structure and inhomoginities
\citep{asp01,gon08}. In addition, \ion{Fe}{i} lines suffer from 
severe NLTE effects in metal-poor stars \citep[see e.g.][]{the99}. 
We note that
\citet{shc05} have performed NLTE computations for the metal-poor
subgiant HD140283 ($[{\rm Fe}/{\rm H}]\sim -2.5$) with a single
snapshot of a 3D hydrodynamical simulation \citep{asp99}, 
and found NLTE-LTE corrections of $+0.9$ and $+0.4$ for \ion{Fe}{i}
and \ion{Fe}{ii} lines, respectively. 

Finally, the oxygen abundance in the Sun is still a matter of debate.
We will adopt throughout this work the value of $\log [N({\rm
O})/N({\rm H}]_\odot =8.76$, which was determined through
3-dimensional (3D) hydrodynamical models \citep{caf08}. 

We have used a subset of the CIFIST
grid of 3D hydrodynamical model atmospheres
\citep{lud09} to investigate the 3D-LTE and 3D-NLTE [O/Fe] and [O/H] trends
in metal-poor dwarf stars.   

\section{3D hydrodynamical simulations}

The  3D hydrodynamical model atmospheres
\citep[see][for further details]{lud09}
were computed with the 
code \cobold\ \citep{fre02,wed04}. Each model consists of a
representative set of snapshots sampling the temporal evolution of the
photospheric flow. In Table~\ref{tabmod} we provide
a summary of the 3D models used in this paper. The evolutionary time
scale and the spatial scale of the convective surface flows roughly
scale inversely proportional to the surface gravity. Following this 
scaling behaviour, we tried to ensure that the duration of each model
sequence corresponds to at least 3600\pun{s} of solar-equivalent time.
Moreover, we also scaled the size of the computational domain leading
to about the same total number of convective cells in each model. 

The spatial resolution was typically 
$(n_\mathrm{x}\times n_\mathrm{y} \times n_\mathrm{z})=(140,140,150)$, 
with a constant grid spacing in the
(horizontal) x- and y-directions and a variable spacing in z-direction. Our
standard solar model -- actually not part of the grid used here, but
mentioned for reference -- has a horizontal
size of 5600~km. The total height is 2250~km, from $z=-1380$~km below optical
depth unity to $z=870$~km above, located in the low chromosphere. The
computational time step was typically 0.1--0.2\pun{s}.

\begin{figure}[!ht]
\centering
\includegraphics[height=8.5cm,angle=90]{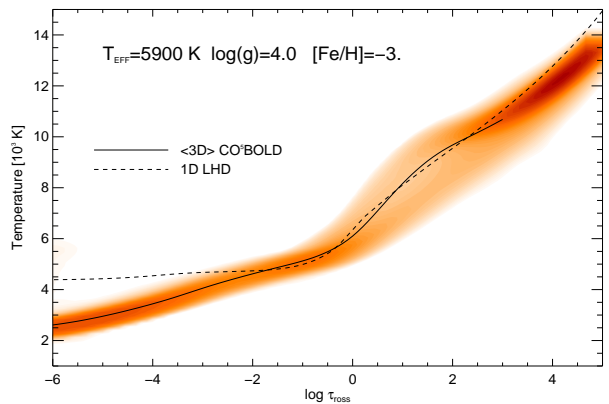}
\caption{3D temperature structure of the 3D \cobold\ atmospheric model,
compared to the average temperature profile,
$\langle\mathrm{3D}\rangle$, and the 1D$_{\rm LHD}$ model. Darker
shades indicate more likely temperatures. The stellar parameters and
metallicity of this 3D simulation are 
$T_{\rm eff}[{\rm K}]/\log(g[{\rm cm~s}^{-2}])/[{\rm Fe}/{\rm H}]=5850/4/-3$.} 
\label{temp3D}
\end{figure}  

\begin{figure}[!ht]
\centering
\includegraphics[height=8.5cm,angle=90]{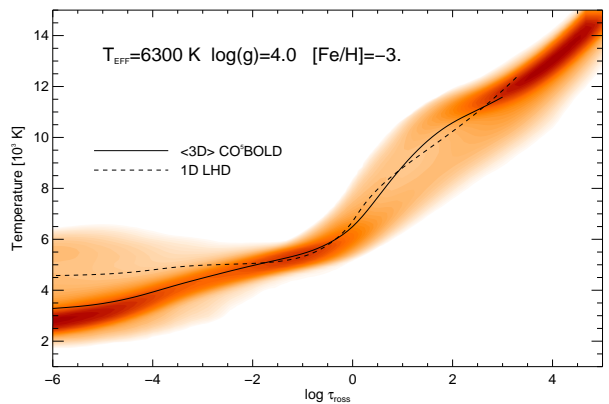}
\caption{Same as Fig.~\ref{temp3D} but for the 3D model 
$T_{\rm eff}[{\rm K}]/\log(g[{\rm cm~s}^{-2}])/[{\rm Fe}/{\rm H}]=6270/4/-3$.}   
\label{temp3Db}
\end{figure}  

The description of the radiative energy exchange is important for the 
resulting temperature structure of a model. Here we provide some
details about the binning-scheme which we applied when modelling the
wavelength dependence of the radiative transfer. The wavelength
dependence is represented by 5 multi-group bins in solar metallicity
models and 6 bins at sub-solar metallicities following the 
procedure originally laid out by \citet{Nordlund82} and subsequently
refined by \citet{Ludwig92,Ludwig+al94,Voegler+al04}. 
For test purposes we calculated a few models with 12 bins.
Different from the original implementation of \citet{Nordlund82}, 
the bin-averaged opacities were calculated explicitely from
the run of the monochromatic opacity within a particular bin. No
scaling among the bins was assumed. The sorting
into wavelength groups was done applying thresholds in logarithmic
Rosseland optical depth $\{ +\infty, 0.0, -1.5, $ $-3.0, -4.5,
-\infty\}$ for the 5-bin, and $\{+\infty, 0.1, 0.0,$ $-1.0, -2.0,
-3.0, -\infty\}$ for the 6-bin schemes. For the 12 bins we used
as thresholds $\{ +\infty, 0.15, 0.0, $ $ -0.75, -1.5, -2.25, $ $-3.0,
-3.75, -4.5, -\infty\}$; in addition, each of the first three
continuum-like bins were split into 2 bins according to wavelength at
550, 600, and 650\,nm. In all but one bin a switching between
Rosseland and Planck averages was performed at a band-averaged
Rosseland optical depth of 0.35; in the bin gathering the largest line
opacities the Rosseland mean opacity was used throughout. The
decisions about the number of bins, and sorting thresholds are motivated
by comparing radiative fluxes and heating rates obtained by the binned
opacities in comparison to high wavelength resolution. In
comparison to the models of \citet{asp01}, who worked with
Stein-Nordlund models \citep{ste98} and used 4 opacity bins, 
our treatment of the wavelength dependence of the radiation field 
puts extra emphasis on the continuum-forming layers. 

\begin{table}[!ht]
\caption[]{Details of the 3D hydrodynamical \cobold\ model atmospheres.}  
\label{tabmod}
\centering
\begin{tabular}{lrrrr}
\hline
\hline
\noalign{\smallskip}
$\langle T_{\rm eff} \rangle$$^a$ & $\log{g}$ & [Fe/H] & Time$^b$ & Snapshots \\       
\noalign{\smallskip}
 [K] & [cgs] & [dex] & [s] &  \\       
\noalign{\smallskip}
\hline
\noalign{\smallskip}
4920 & 3.5 &  0 & 132000  & 19 \\
4930 & 3.5 & --1 & 420000  & 20 \\
4980 & 3.5 & --2 & 544000  & 20 \\
4980 & 3.5 & --3 & 464000  & 20 \\
4950 & 4.0 &  0 & 44000   & 20 \\
4990 & 4.0 & --1 & 235200  & 20 \\
4960 & 4.0 & --2 & 58401   & 20 \\
4990 & 4.0 & --3 & 54400   & 20 \\
4990 & 4.5 &  0 & 31200   & 20 \\
5060 & 4.5 & --1 & 56500   & 19 \\
5010 & 4.5 & --2 & 46500   & 19 \\
4990 & 4.5 & --3 & 29700   & 20 \\
5430 & 3.5 &  0 & 144000  & 18 \\
5480 & 3.5 & --1 & 106800  & 19 \\
5500 & 3.5 & --2 & 46800   & 20 \\
5540 & 3.5 & --3 & 201600  & 20 \\
5480 & 4.0 &  0 & 34800   & 20 \\
5530 & 4.0 & --1 & 95200   & 20 \\
5470 & 4.0 & --2 & 33800   & 20 \\
5480 & 4.0 & --3 & 60800   & 20 \\
5490 & 4.5 &  0 & 19000   & 20 \\
5470 & 4.5 & --1 & 48600   & 20 \\
5480 & 4.5 & --2 & 57000   & 20 \\
5490 & 4.5 & --3 & 54300   & 20 \\
5880 & 3.5 &  0 & 236000  & 20 \\
5890 & 3.5 & --1 & 200000  & 20 \\
5860 & 3.5 & --2 & 112000  & 20 \\
5870 & 3.5 & --3 & 112000  & 20 \\
5930 & 4.0 &  0 & 26400   & 18 \\
5850 & 4.0 & --1 & 27000   & 20 \\
5860 & 4.0 & --2 & 30000   & 20 \\
5850 & 4.0 & --3 & 30500   & 20 \\
5870 & 4.5 &  0 & 16900   & 19 \\
5920 & 4.5 & --1 & 7000    &  8 \\
5920 & 4.5 & --2 & 24500   & 18 \\
5920 & 4.5 & --3 & 9000    & 19 \\
6230 & 4.0 &  0 & 54000   & 20 \\
6260 & 4.0 & --1 & 43800   & 20 \\
6280 & 4.0 & --2 & 27600   & 16 \\
6270 & 4.0 & --3 & 46200   & 20 \\
6230 & 4.5 &  0 & 35400   & 20 \\
6240 & 4.5 & --1 & 25000   & 20 \\
6320 & 4.5 & --2 & 9100    & 19 \\
6270 & 4.5 & --3 & 22000   & 18 \\
6490 & 4.0 &  0 & 58200   & 20 \\
6500 & 4.0 & --1 & 57000   & 20 \\
6530 & 4.0 & --2 & 49200   & 20 \\
6410 & 4.0 & --3 & 56400   & 20 \\
6460 & 4.5 &  0 & 17400   & 20 \\
6460 & 4.5 & --1 & 36200   & 19 \\
6530 & 4.5 & --2 & 9100    & 19 \\
6450 & 4.5 & --3 & 2200    & 12 \\
\noalign{\smallskip}
\hline     
\end{tabular}
\begin{list}{}{}
\item[$^{a}$] Temporal average of the emergent \teffo.
\item[$^{b}$] Total time covered by the temporal evolution of the
photospheric flow named as snapshots.
\end{list}
\end{table}

\subsection{3D temperature structure\label{sectstr}}

In Figs.~\ref{temp3D} and~\ref{temp3Db} we depict the temperature
structure of two 3D hydrodynamical simulations for two different
effective temperatures, 5850 and 6270\,K, and the same surface gravity,
$\log(g/{\rm cm~s}^2)=4$, and metallicity, $[{\rm Fe}/{\rm H}]=-3$. 
The main difference between these two 3D models is the temperature
inhomogeneities which are larger in the hotter model.
These strong temperature fluctuations are typically present for
effective temperatures hotter than 5900~K and surface gravities lower
than 4. 

\begin{figure}[!ht]
\centering
\includegraphics[width=8.5cm,angle=0]{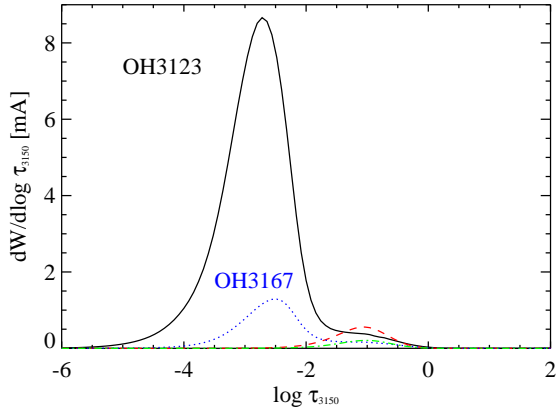}
\caption{Contribution functions to the equivalent width of the OH
3123~{\AA} line computed with \cobold 3D model (solid line) and
1D$_{\rm LHD}$ model (dashed line). The same is displayed for the OH
3167~{\AA} line (3D model, dotted line, and 1D$_{\rm LHD}$ model,
dashed-dotted line).
The stellar parameters and metallicity of this simulation are 
$T_{\rm eff}[{\rm K}]/\log(g[{\rm cm~s}^{-2}])/[{\rm Fe}/{\rm H}]=6270/4/-3$.}   
\label{cf3DOH}
\end{figure}  

\begin{figure}[!ht]
\centering
\includegraphics[width=8.5cm,angle=0]{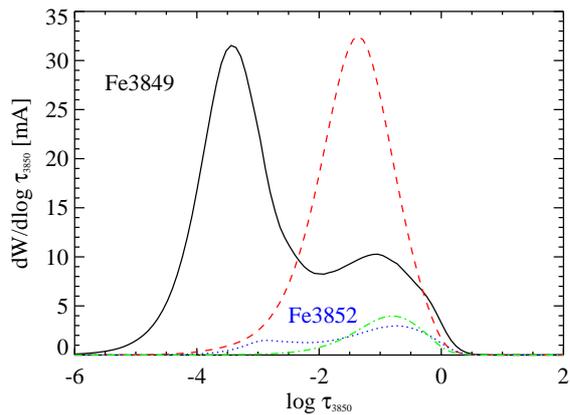}
\caption{Same as Fig.~\ref{cf3DOH} but for the
\ion{Fe}{i} 3849~{\AA} and~3852~{\AA} lines.}   
\label{cf3DFe}
\end{figure}  

\begin{figure}[!ht]
\centering
\includegraphics[width=8.5cm,angle=0]{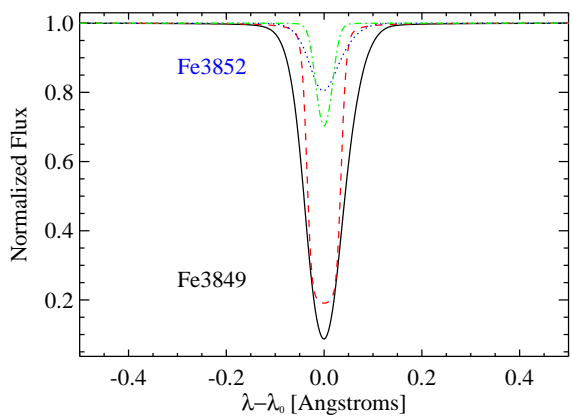}
\caption{Line profiles of the \ion{Fe}{i}~3849~{\AA} and~3852~{\AA}
lines computed with \cobold 3D model (solid line for
\ion{Fe}{i}~3849~{\AA} and dotted line for \ion{Fe}{i}~3852~{\AA}) and
1D$_{\rm LHD}$ model (dashed line for \ion{Fe}{i}~3849~{\AA} and
dashed-dotted line for \ion{Fe}{i}~3852~{\AA}).    
The stellar parameters and metallicity of this simulation are 
$T_{\rm eff}[{\rm K}]/\log(g[{\rm cm~s}^{-2}])/[{\rm Fe}/{\rm H}]=6270/4/-3$.}
\label{profileFe}
\end{figure}  

The 3D models displayed in Figs.~\ref{temp3D} and~\ref{temp3Db} show an
overcooling effect in the outer layers of the star with respect
to 1D models. This is particularly
relevant in all models with metallicities [Fe/H$]=-3$ and~$-2$. 3D
models with metallicities [Fe/H$]=-1$ and~$0$ do not show a pronounced
overcooling effect and the line formation is dominated by the temperature
fluctuations. We note that the overcooling of the higher photospheric
layers is the dominant 3D effect, but in the deep photosphere, at 
$\log \tau \sim -1$, 3D models are always slightly hotter than a
corresponding hydrostatic 1D model in radiative-convective 
equilibrium.

The comparison of 3D versus 1D models depends on which particular 1D model is
chosen. We compared each of our full 3D models to a corresponding (in
\teffo, \loggo, and metallicity) standard hydrostatic 1D model
atmosphere (hereafter denoted as 1D$_{\rm LHD}$) and a 1D model
obtained from the temporal and horizontal average of the 3D structure
over surfaces of equal (Rosseland) optical depth (hereafter denoted as
$\langle$3D$\rangle$). The 1D$_{\rm LHD}$ model is calculated with a 1D
atmosphere code called LHD. It assumes plane-parallel geometry and
employs the same micro-physics (equation-of-state, opacities) as
\cobold. Convection is described by mixing-length theory. See
\citet{caffauS} for further details.  Note that the 1D$_{\rm LHD}$
model depends on the choice of the mixing-length parameter, the
$\langle$3D$\rangle$ model on details of the averaging procedure. 
The temperature of the $\langle$3D$\rangle$ model is indeed obtained by
averaging the fourth power of the 3D temperature field. The choice was
motivated by the property that this kind of averaging largely preserves
the radiative flux \citep{Steffen+al95}. 

We adopted a mixing-length parameter $\alpha=1$ in
the LHD models. However, this parameter only affects lines that are
formed at optical depth $\log \tau > -0.5$. As we will see in the next
section, the OH lines in metal-poor 1D models form at optical depth
$\log \tau < -1$. 

We compute the curves of growth of the OH lines and \ion{Fe}{i} lines,
for both 1D and 3D models with the spectral synthesis code {\linfor}
\footnote{more information on {\linfor} can be found in 
the following link:
http://www.aip.de/$\sim$mst/Linfor3D/linfor\_3D\_manual.pdf}.
It is necessary to assume a value of the micro-turbulence for the 
spectral synthesis of the 1D models. 
We have adopted a \mic~$=1$~\kmso.  

\subsection{Line formation\label{seclfor}}

In Fig.~\ref{cf3DOH} we display the contribution functions (CFs) to
the equivalent width (EW) of the disc centre of the OH lines at 3123~{\AA} 
and~3167~{\AA} according to the \cobold 3D model with given stellar parameters 
and metallicity 
$T_{\rm eff}[{\rm K}]/\log(g[{\rm cm~s}^{-2}])/[{\rm Fe}/{\rm H}]=6270/4/-3$.
These two features have different excitation
potentials ($\chi=0.2$ and~$1.1$ eV for the OH 3123~{\AA} and
3167~{\AA} lines, respectively) and their contribution functions 
look similar in shape but different in strength. 
These molecular lines tend to form in the outer layers of the 3D model
atmosphere, at $\log \tau$ between~$\sim -2.8$ and $\sim-2.2$, where
the 3D model is cooler than the 1D$_{\rm LHD}$ model. Thus 
this line is significantly weaker in the 1D$_{\rm LHD}$ model and 
forms in the inner layers of the star. Because of the sensitivity 
of OH lines to the temperature, the line appears much stronger 
in 3D than in 1D.  

\begin{figure}[!ht]
\centering
\includegraphics[width=8.5cm,angle=0]{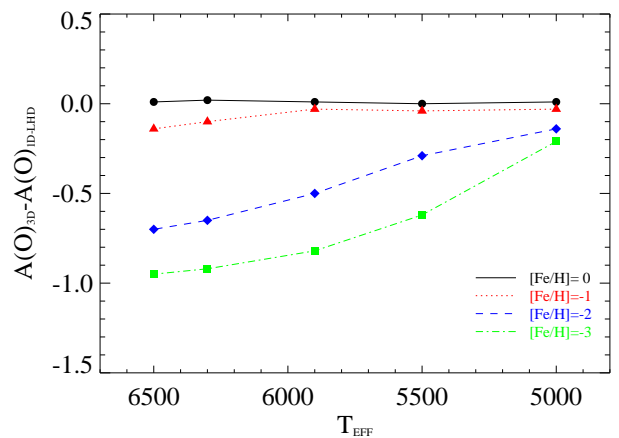}
\caption{{\ftac} corrections for different \teff and metallicities,
computed for the OH 3167~{\AA} line. These 3D hydrodynamical model
atmospheres have \logg$=4$.}   
\label{ct3D}
\end{figure}  

\begin{figure}[!ht]
\centering
\includegraphics[width=8.5cm,angle=0]{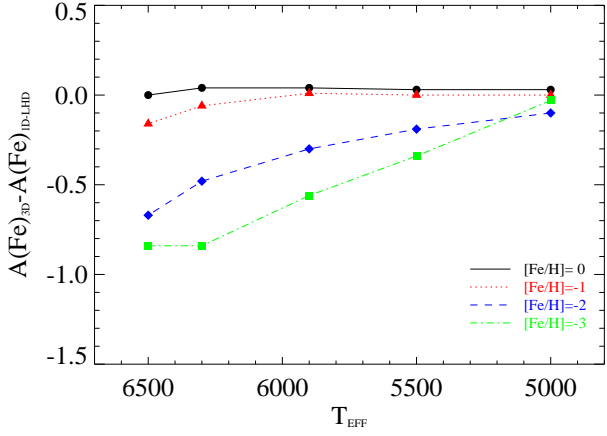}
\caption{Same as Fig.~\ref{ct3D} but for the \ion{Fe}{i} 3849~{\AA}.}  
\label{ct3DFe1}
\end{figure}  

\begin{figure}[!ht]
\centering
\includegraphics[width=8.5cm,angle=0]{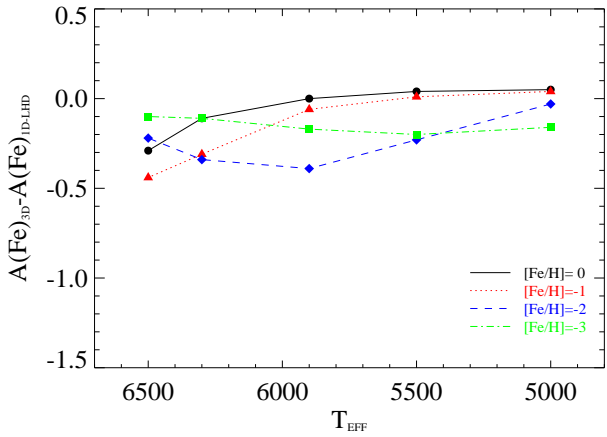}
\caption{Same as Fig.~\ref{ct3D} but for the \ion{Fe}{i} 3852~{\AA}.}  
\label{ct3DFe2}
\end{figure}  

\begin{figure}[!ht]
\centering
\includegraphics[width=8.5cm,angle=0]{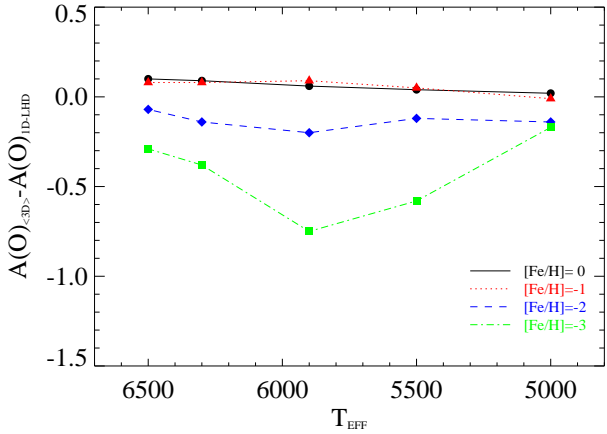}
\caption{{\foac} corrections for different \teff and metallicities,
computed for the OH 3167~{\AA} line. These 3D hydrodynamical model
atmospheres have \logg$=4$.}   
\label{ct1D}
\end{figure}  

This effect is usually found in metal-poor atmospheres
\citep[see][]{asp99} and is particularly important for the oxygen
abundances derived from OH molecules, as the differences are large in
the layers where these lines are formed. The result is that the oxygen
abundances become lower in the 3D formulation than in the 1D. We
quantify this effect through the 3D correction (see
Sect.~\ref{sec3dcor}). In hotter 3D models (\teff$\gtrsim5900$~K) this
effect becomes more severe, making the 3D corrections larger.

We display the CFs to the EW of the \ion{Fe}{i} lines at 3849~{\AA} 
and~3852~{\AA} in Fig.~\ref{cf3DFe}. These lines have different excitation
potentials ($\chi=1.0$ and~$2.2$ eV for the \ion{Fe}{i} 3849~{\AA} and
3852~{\AA} lines) and their CFs are different in both strength and
shape. 
The \ion{Fe}{i} 3849~{\AA} line extends from $\log \tau\sim-4.5$ to
almost 0, whereas the \ion{Fe}{i} 3852~{\AA} line, due to the higher 
excitation potential, goes from $\log \tau\sim-3.5$ to 0. 
Nevertheless, the main contribution is located at $\log \tau\sim-3.5$
for the \ion{Fe}{i} 3849~{\AA} line and at $\log \tau\sim-1$ for the
\ion{Fe}{i} 3852~{\AA} line. 
This means that the \ion{Fe}{i} 3849 line is mostly formed 
in the outer layers of the 3D atmospheric model, 
and the equivalent width is larger than in the 1D$_{\rm LHD}$ model. 
Although the \ion{Fe}{i} 3852~{\AA} line is mainly formed 
in the inner layers, the EW computed for this 3D model is still 
larger than in 1D, due to the contribution of the outer layers. 
Therefore, this line still provides negative 3D corrections.  
In addition, although they have very similar excitation potentials, the
ratio EW(3D)/EW(1D$_{\rm LHD}$) of the \ion{Fe}{i} 3849~{\AA} line is
smaller than that of the OH 3167~{\AA} line. This is probably caused
by the stronger sensitivity of the OH line to the temperature 
structure of the 3D model. 

In Fig.~\ref{profileFe} we display the line profiles of the
\ion{Fe}{i} 3849~{\AA} and~3852~{\AA} lines for the same 3D and
1D$_{\rm LHD}$ models with stellar parameters and metallicity 
$T_{\rm eff}/\log(g/{\rm cm~s}^2)/[{\rm Fe}/{\rm H}]=6270/4/-3$. 
The 3D profiles of both lines are more broadened than the
1D$_{\rm LHD}$ profiles in part because these 1D profiles were not
been convolved with a macroturbulent velocity, whereas
in the 3D profiles the broadening effect due to macroturbulence is
naturally included.
In addition, the 3D profiles are slightly
asymmetric due to the convective motions in the hydrodynamical
simulations included in the 3D model atmosphere. It is quite clear
that the 3D profile of the \ion{Fe}{i} 3849~{\AA} line is stronger
than the 1D$_{\rm LHD}$ profile. This is not as evident for the
\ion{Fe}{i} 3852~{\AA} line, where the line core is deeper in the
1D$_{\rm LHD}$ profile but the line wings are tighter. The ratios
EW(3D)/EW(1D$_{\rm LHD}$) of these lines in these 3D and 1D$_{\rm
LHD}$ models are 1.59 and 1.25 for the \ion{Fe}{i} 3849~{\AA}
and~3852~{\AA} lines, respectively. As a comparison, the ratios
EW(3D)/EW(1D$_{\rm LHD}$) of the OH 3123~{\AA} and~3167~{\AA} lines
are 20.2 and 8.2, respectively. Therefore, it is expected that all
these lines should provide different 3D corrections. 
We will discuss quantitatively these abundance
corrections in Sect.~\ref{sec3dcor}.
 
\begin{table*}[!ht]
\caption[]{3D abundance corrections for several OH lines using 3D 
\cobold\ models with 6 and 12 opacity bins with the parameters 
$T_{\rm eff}[{\rm K}]/\log(g[{\rm cm~s}^{-2}])/[{\rm Fe}/{\rm H}]=6270/4/-3$ 
and $6240/4/-3$, respectively.}     
\label{tab3dcor}
\centering
\begin{tabular}{lrrrrrrr}
\hline
\hline
\noalign{\smallskip}
Wavelength & $\chi$ & \tac$_{{\rm ,}6-{\rm bin}}$ & \tac$_{{\rm ,}12-{\rm bin}}$ & $\Delta_{6-12{\rm bin}}$ & \ftac$_{{\rm ,}6-{\rm bin}}$ & \ftac$_{{\rm ,}12-{\rm bin}}$ & $\Delta_{6-12{\rm bin}}$\\       
\noalign{\smallskip}
 [\AA] & [eV] & [dex] & [dex] & [dex] & [dex] & [dex] & [dex]\\       
\noalign{\smallskip}
\hline
\noalign{\smallskip}
3128.287 & 0.210 & -0.852 & -0.953 & 0.101 & -1.416 & -1.057 & -0.359 \\
3139.170 & 0.763 & -0.647 & -0.735 & 0.088 & -1.083 & -0.806 & -0.277 \\
3167.169 & 1.108 & -0.544 & -0.622 & 0.078 & -0.910 & -0.675 & -0.235 \\
\noalign{\smallskip}
\hline     
\end{tabular}
\end{table*}

\subsection{3D corrections\label{sec3dcor}}
     
The spectral synthesis code {\linfor} produces three curves of growth:
full 3D, $\langle$3D$\rangle$ and 1D$_{\rm LHD}$ \citep{caf09,caf10}. 
This allows us to estimate abundance corrections through 
the EW(3D) of the 
3D model abundance to derive the corresponding abundances in the
$\langle$3D$\rangle$ and 1D$_{\rm LHD}$ models.
Two main effects distinguish 3D from 1D models: the
average temperature profile and the horizontal temperature
fluctuations. We quantify the contribution of these two main effects 
by introducing the 3D correction as {\ftac}. The first and the
second effect can also be separately estimated with the {\foac}
and {\tac} corrections, respectively. The velocity fluctuations 
 (i.e. intrinsic velocity fluctuations and microturbulence) are other
effects that also play a role, but their influence on the derived
abundance, although not negligible, is not as relevant.
In Fig.~\ref{ct3D} we display the {\ftac} abundance corrections of the
OH 3167~{\AA} line for different effective temperatures and
metallicities, and \logg$=4$. The 3D models with metallicities
[Fe/H$]=-1$ and~0 show very small and mainly positive {\ftac}
corrections, whereas 3D models with [Fe/H$]=-3$ and~-2 show strong,
negative corrections.  
The {\ftac} generally increases towards hotter 3D models and lower 
metallicites.
We note here that the behaviour of other OH lines is practically the
same although for slightly different values for the {\ftac} abundance
corrections. 

In Figs.~\ref{ct3DFe1} and~\ref{ct3DFe2} we show the {\ftac} abundance
corrections of the \ion{Fe}{i} 3849~{\AA} and~3852~{\AA} lines 
for different effective temperatures and metallicities, and \logg$=4$. 
The behaviour of the \ion{Fe}{i} 3849~{\AA} line is quite similar to
that of the OH 3167~{\AA} line, although with smaller {\ftac} values.
On the other hand, the \ion{Fe}{i} 3852~{\AA} line behaves in a completely 
different way, and the {\ftac} corrections depend on whether the
line formation occurs mainly in the outer or inner region of the
atmosphere of a given 3D model at a given temperature and metallicity.

In Fig.~\ref{ct1D} we depict the {\foac} corrections of the OH
3167~{\AA} line for different effective temperatures and
metallicities, and \logg$=4$. The {\foac} corrections reflect 
how important these cooling effects are in the 3D models. Thus, 
the strongest correction is found for the 3D model with metallicity
[Fe/H$]=-3$ and \teff$\sim\, 5900$. One also realizes by comparing
Figs.~\ref{ct3D} and~\ref{ct1D} that the 3D models with
metallicities [Fe/H$]=-3$ and~--2 and \teff$>\, 5900$ have large
temperature fluctuations that account for a significant fraction of
the total {\ftac} correction.

\begin{figure}[!ht]
\centering
\includegraphics[width=8.5cm,angle=0]{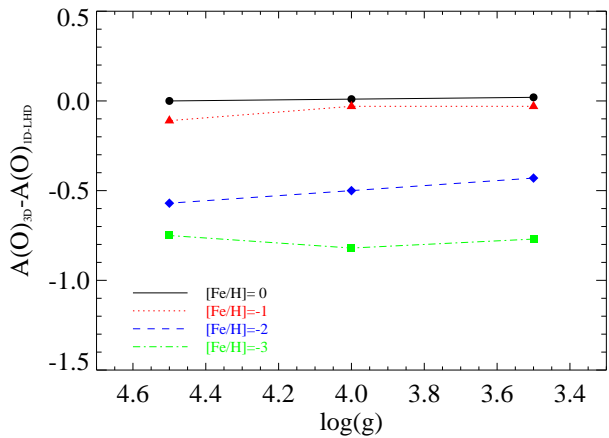}
\caption{{\ftac} corrections for different surface gravities and
metallicities, computed for the OH 3167~{\AA} line. These 3D
hydrodynamical model atmospheres have \teff$\sim5900$~K.}  
\label{cg3D}
\end{figure}

The sensitivity of the {\ftac} corrections to the surface gravity is
less relevant (see Fig.~\ref{cg3D}). The derived {\ftac} corrections
are typically larger in 3D models with larger surface gravities,
although only variations of $\lesssim 0.2$~dex are expected from
\logg$=3.5$ to 4.5.

\begin{figure}[!ht]
\centering
\includegraphics[width=8.5cm,angle=0]{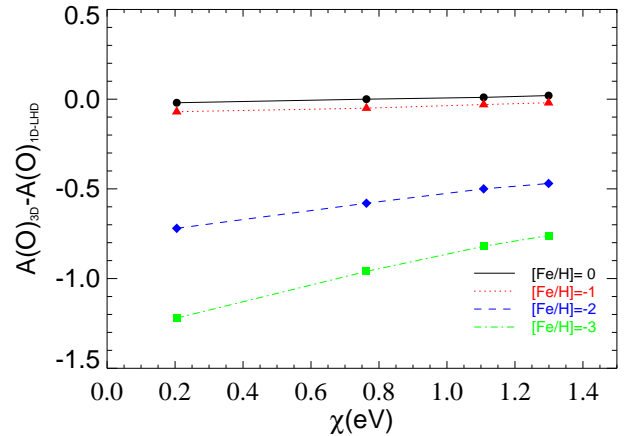}
\caption{{\ftac} corrections for different excitation potentials and
metallicities, computed for OH lines. These 3D hydrodynamical models
have \teff$\sim5900$~K and \logg$=4$.}  
\label{cep3D}
\end{figure}  

The excitation potential determines the layer of the 3D model
atmosphere where the line forms. Thus, it is expected that a line of
the given molecule or atom forms further out if the excitation
potential is lower. In the case of OH molecular lines, further out
means lower temperature and therefore enhanced molecule formation.
This is exactly what it is displayed in Fig.~\ref{cep3D}, the lower the
excitation potential of a OH line is, the larger negative {\ftac}
corrections it provides.

All in all, the dependence of the abundance corrections on
metallicity shows a kind of threshold behaviour. Noticeable corrections
indicating sizable temperature differences between 1D and 3D models
set in rather abruptly at metallicities below --1. \citet{lud08}
have argued that this can be expected on general grounds emerging from
the temperature and metallicity sensitivity of the radiative
relaxation time scale.

\subsection{Dependence on carbon abundance\label{seccarbon}}
 
The molecular equilibrium of OH lines also involves the molecules CO,
CH and C$_2$. Thus, the C abundance surely affects the contribution 
function to the EW of the OH lines.
The CO molecule is the most tightly bound and has the tendency to 
get hold of all available O atoms, to the detriment of OH.
\citet{beh10} have already checked this effect in C-enhanced 
metal-poor stars, showing that if the C abundance is large enough, the
{\ftac} abundance corrections become much smaller, reaching values
of the order of~$-0.2$ if the carbon-to-oxygen ratio is as high as
[C/O]~$\sim 1$.  In Table~\ref{tab3dco} we provide the {\ftac}
corrections to the O abundance derived from an OH line with excitation 
potential $\sim0.1$~eV, for several values of  [C/O], 
which have been computed only 
for testing purposes with the 3D model with parameters 
$T_{\rm eff}[{\rm K}]/\log(g[{\rm cm~s}^{-2}])/[{\rm
Fe}/{\rm H}]=6270/4/-3$. For these low ratios [C/O], the {\ftac} 
corrections are similar and within 0.15~dex. In addition, looking at
these corrections in detail one realizes that (i) when [C/O] is low
enough, the {\ftac} correction to the O abundance derived from OH
lines is only slightly dependent on the C 
abundance with changes $\lesssim 0.05$ dex, (ii) when the [C/O] is
near the value zero, the C abundance starts to become relevant,
with changes $\sim 0.20$ dex, (iii) only when [C/O] is large
enough, the {\ftac} correction becomes negligible.

In our computations we assumed that the carbon abundance scales with
metallicity, [C/Fe]~$=0$. In models with metallicities lower than
$-1$~dex, we have assumed the $\alpha$-elements enhanced 
by a factor of [$\alpha$/Fe]~$=0.4$. 
Therefore, the [C/O]~$=-0.4$, which means that
at the lowest metallicities, i.e. at [Fe/H]~$=-2$ and~$-3$, where the
observed abundances can reach values of [O/Fe]~$\sim 0.5-1.0$
(see Sect.~\ref{secdisc}) and [C/Fe]~$\sim 0$ \citep[e.g.][]{lai07},
our estimated {\ftac} corrections might have to be corrected downwards
by $0.06-0.12$~dex.  This small effect has not been taken into
account and goes in the direction of slightly increasing the 
strength of the {\ftac} corrections. 

\begin{table}[!ht]
\caption[]{{\ftac} O abundance corrections for several [C/O] ratios
from the OH line 3106~{\AA} with an excitation potential equal to
0.1~eV, using the 3D model with the parameters 
$T_{\rm eff}[{\rm K}]/\log(g[{\rm cm~s}^{-2}])/[{\rm Fe}/{\rm H}]=6270/4/-3$.}     
\label{tab3dco}
\centering
\begin{tabular}{lrrrrrrr}
\hline
\hline
\noalign{\smallskip}
[C/O] & [C/Fe] & [O/Fe] & \ftac \\       
\noalign{\smallskip}
[dex] & [dex] & [dex] & [dex] \\       
\noalign{\smallskip}
\hline
\noalign{\smallskip}
-1.0 & 0.0 & 1.0 & -1.53  \\
-0.6 & 0.4 & 1.0 & -1.49  \\
-0.2 & 0.8 & 1.0 & -1.37  \\
\noalign{\smallskip}
\hline     
\end{tabular}
\end{table}

\subsection{Dependence on opacity binning\label{secopa}}

We tested a few 3D hydrodynamical models
with 12 opacity bins and found that the cooling effect at
$\log \tau \lesssim -2$ is less pronounced than in our standard 6 bin
models. From this result we expect that in the next generation of 3D
models, using a refined opacity binning, the {\ftac} abundance 
corrections would come out smaller for OH lines and Fe lines with 
excitation potentials below 1.5~eV. Therefore, the corrections 
presented here for these lines can be considered as upper limits. 

In Table~\ref{tab3dcor} we provide a comparison between the {\ftac}
corrections in the 6-bin and 12-bin 3D models. The 6-bin model produces
larger corrections for these three OH lines, 
between $\sim -0.36$ and~$-0.23$~dex, 
with excitation potentials between 0.2 and 
1.1~eV. In addition, the 12-bin model has stronger temperature 
fluctuations than the 6-bin model, since the {\tac} is larger for 
the 12-bin model. This difference explains some of the larger scatter 
in oxygen abundances obtained from different lines when using 3D 
models than in the 1D case (see Sect.~\ref{secdisc}).

\section{Applications to the Galactic oxygen abundance\label{secdisc}}

\begin{figure}[!ht]
\centering
\includegraphics[width=8.5cm,angle=0]{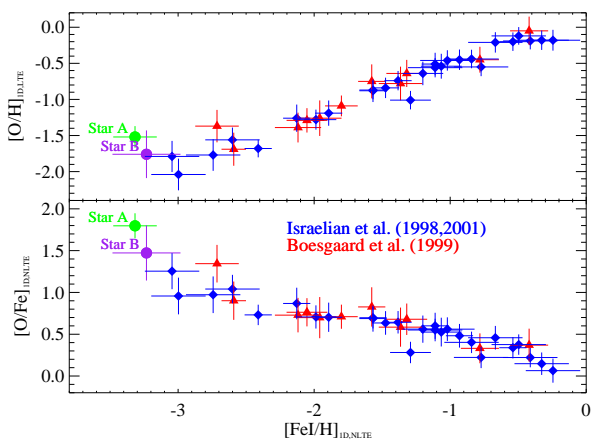}
\caption{1D [O/\ion{Fe}{i}] ratio versus 1D \ion{Fe}{i} abundances in
metal-poor dwarf stars computed with 1D-LTE models using OH lines from
Boesgaard et al.(1999, triangles), Israelian et al.(1998, 2001,
diamonds). The circles represent the most metal-poor dwarf stars of
the binary \mbox{CS 22876--032} from \citet{gon08}. The stars labeled
``A'' and ``B'' represent the primary and secondary star in this
binary.}   
\label{ofe1D}
\end{figure}

The aim of this study is to apply our grid of the {\ftac} corrections
to a sample of metal-poor dwarf stars with available O abundances
derived from near-UV OH lines. 
This sample was obtained from the studies of
\citet{boe99}, \citet{isr98,isr01}, and the most metal-poor dwarf stars
of the binary \mbox{CS 22876--032} from \citet{gon08}. 
We adopted the stellar parameters and the O and \ion{Fe}{i}
abundances of the dwarf stars provided in those works.
In Fig.~\ref{ofe1D} we display the 1D-LTE [O/H] and 1D-NLTE
[O/\ion{Fe}{i}] ratios versus 1D-NLTE [\ion{Fe}{i}/H] in metal-poor
dwarf stars. 

\begin{table}[!ht]
\caption[]{Selected lines for the abundance analysis.}  
\label{tabline}
\centering                          
\begin{tabular}{llrr}
\hline
\hline
\noalign{\smallskip}
Wavelength & Transition & $\chi$ & $\log gf$ \\ 
\noalign{\smallskip}
 [{\AA}] &  & [eV] &  \\       
\noalign{\smallskip}
\hline
\noalign{\smallskip}
3123.949 & OH $P_{11}(9.5)$ & 0.205 & -1.960 \\
3128.287 & OH $P_{22}(8.5)$ & 0.210 & -2.021 \\
3139.170 & OH $Q_{22}(17.5)$ & 0.763 & -1.559 \\
3167.169 & OH $QR_{22}(21.5)$ & 1.108 & -1.541 \\
3255.493 & OH $P_{22}(23.5)$ & 1.299 & -1.809 \\
3843.257 & \ion{Fe}{i} & 3.047 & -0.241  \\
3849.967 & \ion{Fe}{i} & 1.011 & -0.871 \\
3852.573 & \ion{Fe}{i} & 2.176 & -1.185 \\
\noalign{\smallskip}
\hline     
\end{tabular}
\end{table}

We applied the same formula given in \citet{isr01} to correct
the 1D \ion{Fe}{i} abundances for NLTE. This formula was 
derived by these authors by fiting the 1D NLTE-LTE corrections
provided by \citet{the99} as a function of metallicity. 
They also noted that these 
NLTE corrections were not significantly sensitive to the stellar
parameters in their sample of metal-poor stars. This formula provides
corrections of +0.25 and +0.37 at [Fe/H]~$\sim -2$ and $-4$,
respectively. In the literature
there are other studies of the NLTE effect on Fe lines
\citep[e.g.][]{kor03,shc05,mas10} in metal-poor stars, but 
none of them provides a
similar table with NLTE-LTE corrections for a grid of models at
different metallicities.
NLTE effects on the OH lines may be also significant. \citet{asp01}
tried to investigate this effect using a two-level OH molecule and
found NLTE corrections up to +0.25~dex at metallicities below -2. We
note here that in this study we have not taken into account any
possible NLTE effect on the OH lines.
The trend displayed in Fig.~\ref{ofe1D} clearly shows an increasing
[O/\ion{Fe}{i}] ratio towards lower metallicities. The error bars in
this figure only show the dispersion of the abundance measurements
using 1D models. We derived the slope of a linear fit to
all data points shown in the lower panel in Fig.~\ref{ofe1D},
providing $-0.39 \pm 0.03$. This value may be compared the slope 
of the 1D-NLTE [O/Fe] ratio reported in \citet{isr01}, whose
value is $-0.33 \pm 0.02$. The difference is likely related to the
fact that \citet{isr01} used data points from \citet{isr98,isr01}, 
\citet{boe99} and \citet{edv93}.

\begin{figure}[!ht]
\centering
\includegraphics[width=8.5cm,angle=0]{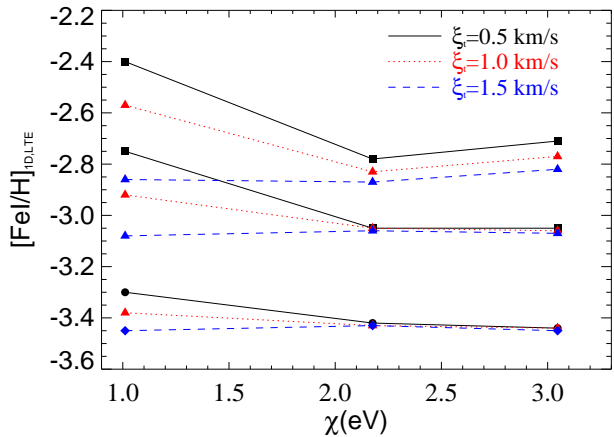}
\caption{1D LTE [Fe/H] abundances of some the stars in Table~\ref{tabafe} excitation
potential of the \ion{Fe}{i} line, for three different microturbulent
velocities adopted in each computation.}  
\label{fehvt}
\end{figure}

We determined for all 3D models in Table~\ref{tabmod} the {\ftac}
abundance corrections for the OH and \ion{Fe}{i} lines given in
Table~\ref{tabline}. The \ion{Fe}{i} lines have three different
excitation potentials from roughly 1 to 3~eV. 
We chose some stars in the sample of \citet{isr98,isr01},
for which we have near-UV spectra to check if the selected \ion{Fe}{i}
lines provide appropriate [Fe/H] values. For this estimate we used the
code 1D LTE MOOG \citep{sne73} and LTE model atmospheres
with $\alpha-$elements enhanced, by +0.4 dex, computed
with the Linux version \citep{sbo04} of the ATLAS code
\citep{kur93}. We used the new opacity distribution functions (ODFs)
of \citet{cak03} with the corresponding metallicity.
In Table~\ref{tabafe} we give
the 1D LTE Fe abundances from these \ion{Fe}{i} lines. The \ion{Fe}{i}
3849~{\AA} line is quite strong and starts to be in the saturation
regime for metallicities above -2. 
If one takes into account that the 3D-LTE corrections for \ion{Fe}{i}
lines of different excitation potentials are different, one should 
obtain different abundances when using 1D models, according 
to Fig.~\ref{cep3D}. We hardly see this strong effect in the values 
given in Table~\ref{tabafe}. In Fig.~\ref{fehvt} we depict these 
\ion{Fe}{i} abundances for different values of microturbulent 
velocities for three of the stars in Table~\ref{tabafe}. 
As soon as one goes to higher metallicities and lower excitation 
potentials, the
\ion{Fe}{i} is more sensitive to the adopted microturbulence. However,
at the lowest metallicities, the \ion{Fe}{i} line at 3~eV provides
more uncertain results because the EW is very low and therefore more
sensitive to the quality of the observed spectra. We believe that our
choice of microturbulent velocity equal to 1~\kms is a reasonable
assumption, although the hottest stars, with \teff$\sim6300$~K,
in our sample may require a slightly higher microturbulence.

In order to derive 3D abundances, we interpolated within the grid of 
{\ftac} corrections using the stellar parameters and
metallicities of the sample stars. 
For OH lines, we applied individual abundance
{\ftac} corrections to the O abundance derived from each OH line and
finally computed the average 3D O abundance of all available lines in
each star. The error of the [O/H] abundance is then estimated from the
dispersion of the 3D-LTE abundances of all OH lines.

\begin{figure}[!ht]
\centering
\includegraphics[width=8.5cm,angle=0]{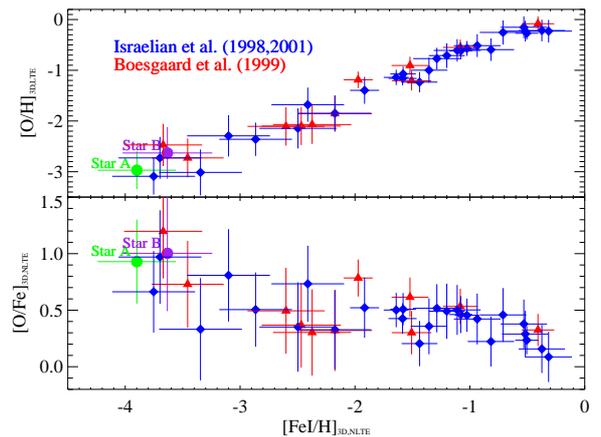}
\caption{3D-LTE [O/H] and 3D-NLTE [O/\ion{Fe}{i}] ratios versus 
3D-NLTE \ion{Fe}{i} abundances computed with 3D-LTE models. The
symbols are the same as Fig.~\ref{ofe1D}.}      
\label{ofe3D}
\end{figure}  

For \ion{Fe}{i} we used a different recipe. 
The papers by \citet{isr98} and \citet{boe99} do
not provide the \ion{Fe}{i} lines used to determine the
metallicity. On the other hand, the paper by \citet{isr01} 
gives a few \ion{Fe}{i} lines in the near-UV spectral region 
where the OH lines form. 
The excitation potential of these \ion{Fe}{i} lines is
$\chi\sim1$~eV. However, these \ion{Fe}{i} line are too strong 
and gets saturated in stars with [Fe/H$]>-2$. 
Therefore, we decided to adopt the following strategy:
(i) for stars with [Fe/H$]<-2$ we only used the \ion{Fe}{i} 3849~{\AA}
line to apply a {\ftac} correction to the 1D metallicities; and (ii)
for stars with [Fe/H$]>-2$ we derived an average correction with the
two other \ion{Fe}{i} lines at 3843 and 3852~{\AA}. For [Fe/H$]>-2$,
we estimated the error of the 3D-LTE [\ion{Fe}{i}/H] by adding
quadratically the dispersion of the {\ftac} corrections and the 
dispersion of the 1D-LTE abundances. For [Fe/H$]<-2$,
we adopted ad hoc a dispersion of 0.3~dex in the {\ftac} abundance 
corrections because we were only using one \ion{Fe}{i} line.
In Sect.~\ref{sec3dcor} we discussed the expected differences in the
3D-LTE {\ftac} correction between the \ion{Fe}{i} 3849~{\AA} and
3852~{\AA} lines for models with [Fe/H$]<-2$. 
Thus, it is clear that if one would use both \ion{Fe}{i} lines 
to estimate the {\ftac} correction for stars with [Fe/H$]<-2$, 
one would get smaller corrections, but the error bar associated 
to these corrections would be very large.  

As a first approximation, we corrected our values 
for the 3D-LTE \ion{Fe}{i} abundances for NLTE effects with
the 1D-NLTE corrections already applied before to the 1D case. 
Finally, we determined the 3D-NLTE [O/\ion{Fe}{i}] ratio by 
computing the difference between the 3D-LTE O abundance and the 
3D-NLTE \ion{Fe}{i} abundance in all dwarf stars of the sample. 
We are aware that these NLTE corrections may be too small,
but we also know that our 3D-LTE corrections in both OH and Fe lines
may be also overestimated because of the opacity binning used to compute
our grid of 3D models (see Sect.~\ref{secopa}). In addition, the
carbon abundance may be another factor to take into account (see
Sect.~\ref{seccarbon}). We see that these effects tend to go in
opposite directions when trying to define the slope, if it exists, 
of the ratio [O/Fe] versus [Fe/H] at the lowest metallicities.

\begin{table*}
\caption[]{1D Fe abundances for six stars in the sample of
\citet{isr98,isr01}, from three \ion{Fe}{i} lines of different 
excitation potentials} 
\label{tabafe}
\centering
\begin{tabular}{lrrrrrrr}
\hline
\hline
\noalign{\smallskip}
Star & \teff & \logg & $\xi_t$ & [Fe/H] & [\ion{Fe}{i}/H]$_{3143}$ & [\ion{Fe}{i}/H]$_{3149}$ & [\ion{Fe}{i}/H]$_{3152}$ \\       
\noalign{\smallskip}
     &  [K]  & [dex] & [\kms]  & [dex] & [dex] & [dex] & [dex] \\       
\noalign{\smallskip}
\hline
\noalign{\smallskip}
G 64-12   & 6318 & 4.20 & 1.0 & -3.37 & -3.44 & -3.38 & -3.43 \\
G 64-37   & 6310 & 4.20 & 1.0 & -3.22 & -3.22 & -3.21 & -3.19 \\
LP 815-43 & 6265 & 4.54 & 1.0 & -3.05 & -3.06 & -2.92 & -3.05 \\
HD 140283 & 5550 & 3.35 & 1.0 & -2.70 & -2.77 & -2.57 & -2.83 \\
HD 166913 & 5965 & 3.90 & 1.0 & -1.80 & -- & -1.57 & -1.78 \\
HD 76932  & 5800 & 3.85 & 1.0 & -1.10 & -- & -1.48 & -1.45 \\
\noalign{\smallskip}
\hline     
\end{tabular}
\end{table*}

In Fig.~\ref{ofe3D} we depict the 3D-LTE [O/H] and 3D-NLTE
[O/\ion{Fe}{i}] trend in metal-poor dwarf stars. 
The slope of the linear fit to the 3D-NLTE [O/\ion{Fe}{i}] trend
is smaller in absolute value, $-0.16 \pm0.04$, but still negative and
relatively significant, at least under the assumptions made 
in this work.
The error bars are larger in this figure than in Fig~\ref{ofe1D}
because we added quadratically the standard deviation 
on the OH and Fe abundances, and the dispersion on the 
{\ftac} abundance corrections from different OH and Fe lines. 

\begin{figure}[!ht]
\centering
\includegraphics[width=8.5cm,angle=0]{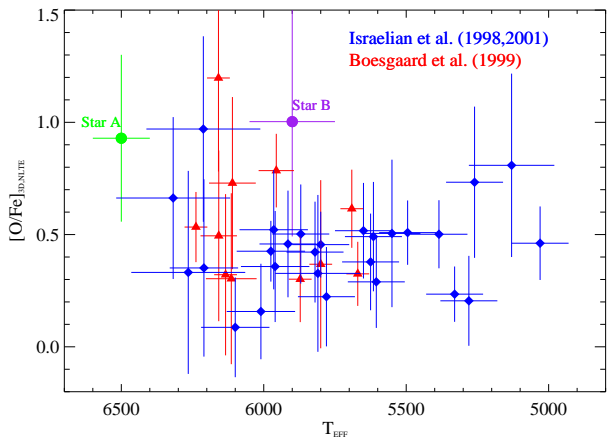}
\caption{3D [O/\ion{Fe}{i}] ratio versus effective temperature. The
symbols are the same as Fig.~\ref{ofe1D}.}   
\label{oteff3D}
\end{figure}

There is also a larger scatter in the almost linear relation
between the 3D-NLTE [O/Fe] and [Fe/H] ratios. 
The reason may be that the 3D corrections applied to the stars 
depend not only on the metallicity of the star, but also on 
the stellar parameters. Therefore, when we
apply the {\ftac} corrections, different stars move to different
directions in the plane [O/Fe] versus [Fe/H].  
As we stated in Sect.~\ref{intro}, large and positive 3D-NLTE
corrections for \ion{Fe}{i} lines are expected in metal-poor 3D 
models \citep{shc05}, but this needs to be done for the whole grid 
of 3D hydrodynamical models to see how the picture changes. We also
note here that the computations done by \citet{shc05} do not consider
the inelastic collisions with neutral hydrogen atoms, because there is
at present no reliable values for the collisional rates, and
the classical Drawin formula leads to uncertain estimates.
They also pointed out that inelastic collisions with neutral hydrogen
atoms would tend to alleviate the NLTE effects, and therefore, the
difference between the NLTE and LTE may be considered as the maximum
effect that inelastic collisions might produce.
Figure~\ref{oteff3D} shows the 3D-NLTE [O/\ion{Fe}{i}] ratio versus
effective temperature of the star. There is no clear trend with
effective temperature, except that due to selection
effects, most of the dwarf stars with higher 3D-NLTE [O/\ion{Fe}{i}]
values are hot stars (\teff$> 5900$), with the exception of very
few cases. 

Finally, we remark here that after all these
corrections, the trend in [O/\ion{Fe}{i}] obtained from near-UV OH lines
in metal-poor dwarf stars is consistent with [O/Fe] ratios obtained
from the forbidden [\ion{O}{i}] in metal-poor giant stars for which the 3D
corrections are expected to be negligible \citep{gon08}.
 
\section{Summary}

The large grid of 3D hydrodynamical model atmospheres of dwarf 
stars quite captivated our attention during the last
three years \citep{lud09}. It took thousands of hours in computing 
time to build such a grid. We have used 52 3D models 
extracted from this grid 
with the following stellar parameters and metallicities: 
\teff$=5000$, 5500, 5900, 6300 and 6500~K, \logg$=3.5$, 4 and 4.5, 
and [Fe/H$]=0$, --1, --2 and --3. The main difference with the 
``classical'' 1D models is that 3D models show temperature 
inhomogeneities and a cooler average
temperature profile in models with [Fe/H$]<-1$.

This allowed us for the first time to compute 3D abundance
corrections of several near-UV OH and \ion{Fe}{i} lines. These
3D corrections are generally larger for higher effective temperatures,
larger surface gravities, and lower metallicities. In addition, lines
with lower excitation potentials show stronger 3D corrections.

We applied this grid of 3D corrections to a sample of metal-poor dwarf
stars from \citet{isr98,isr01}, \citet{boe99} and the most metal-poor
dwarf stars of the binary \mbox{CS 22876--032} from \citet{gon08}. We
interpolated within this grid, using the stellar parameters and
metallicities of these stars. 

Finally, we are able to display a new trend with the 
3D-NLTE [O/\ion{Fe}{i}] ratio versus metallicity and this trend still
increases towards lower metallicities, as the 1D-NLTE, but with a
smaller slope in absolute value. However, we caution 
that some assumptions (as e.g. the restriction to a 6-bin
scheme in most 3D models and NLTE corrections only in 1D), which 
we believe to be reasonable, have been adopted to achieve this result
and that therefore, this result should be taken with that in mind. 
A full 3D-NLTE study of all 3D model atmospheres presented in this 
paper must be performed for
both near-UV OH lines and \ion{Fe}{i} lines to see if this trend
changes. 

\begin{acknowledgements}

J. I. G. H., P. B., H.-G. L. and N. B. acknowledge support from the
EU contract MEXT-CT-2004-014265 (CIFIST). J.I.G.H. also thanks support 
from the project AYA2008-00695 of the Spanish Ministry of Education 
and Science. B.F.\ acknowledges financial support from
the {\sl Agence Nationale de la Recherche} (ANR), and the
{\sl ``Programme National de Physique Stellaire''} (PNPS) of
CNRS/INSU, France. 
We are also grateful to the supercomputing centre CINECA, which 
has granted us time to compute part of the hydrodynamical models 
used in this investigation, through the INAF-CINECA agreement 2006,
2007.

\end{acknowledgements}


\begin{thebibliography}{}

\bibitem[Abia \& Rebolo(1989)]{abi89} 
Abia, C., \& Rebolo, R.\ 1989, \apj, 347, 186 

\bibitem[Asplund et al.(1997)]{asp97}
Asplund, M., Gustafsson, B., Kiselman, D., Eriksson, K. 1997, \aap,
318, 521  

\bibitem[Asplund et al.(1999)]{asp99} 
Asplund, M., Nordlund, {\AA}., Trampedach, R., Stein, R.F.\ 1999, \aap, 346, L17 

\bibitem[Asplund \& Garc{\'{\i}}a P{\'e}rez(2001)]{asp01} 
Asplund, M., \& Garc{\'\i}a P{\'e}rez, A.~E.\ 2001, \aap, 372, 601 

\bibitem[Barbuy(1988)]{bar88} 
Barbuy, B.\ 1988, \aap, 191, 121 

\bibitem[Behara et al.(2010)]{beh10} 
Behara, N.~T., Bonifacio, P., Ludwig, H.~-., Sbordone, L., 
Gonz{\'a}lez Hern{\'a}ndez, J.~I., \& Caffau, E.\ 2010, \aap, in press,
arXiv:1002.1670 

\bibitem[Boesgaard et al.(1999)]{boe99} 
Boesgaard, A.~M., King, J.~R., Deliyannis, C.~P., \& Vogt, S.~S.\
1999, \aj, 117, 492  

\bibitem[Caffau et al.(2007)]{caffauS} 
Caffau, E., Faraggiana, R., Bonifacio, P., Ludwig, H.-G., \& Steffen,
M.\ 2007, \aap, 470, 699  

\bibitem[Caffau et al.(2008)]{caf08} 
Caffau, E., Ludwig, H.-G., Steffen, M., Ayres, T.~R., Bonifacio, P.,
Cayrel, R., Freytag, B., \& Plez, B.\ 2008, \aap, 488, 1031  

\bibitem[Caffau et al.(2009)]{caf09} 
Caffau, E., Ludwig, H.-G., \& Steffen, M.\ 2009, Memorie della Societa 
Astronomica Italiana, 80, 643 

\bibitem[Caffau et al.(2010)]{caf10} 
Caffau, E., Ludwig, H.-G., Steffen, M., Freytag, B., \& Bonifacio, 
P.\ 2010, arXiv:1003.1190 

\bibitem[Castelli \& Kurucz(2003)]{cak03} 
Castelli, F., \& Kurucz, R.~L.\ 2003, Modelling of Stellar
Atmospheres, 210, 20P  

\bibitem[Cayrel et al.(2004)]{cay04}
Cayrel, R., et al.\ 2004, \aap, 416, 1117 

\bibitem[Edvardsson et al.(1993)]{edv93} 
Edvardsson, B., Andersen, J., Gustafsson, B., Lambert, 
D.~L., Nissen, P.~E., \& Tomkin, J.\ 1993, \aap, 275, 101 


\bibitem[Freytag et al.(2002)]{fre02} 
Freytag, B., Steffen, M., \& Dorch, B.\ 2002, Astronomische
Nachrichten, 323, 213  

\bibitem[Fulbright \& Johnson(2003)]{ful03} 
Fulbright, J.~P., \& Johnson, J.~A.\ 2003, \apj, 595, 1154 

\bibitem[Garc{\'\i}a P\'erez et al.(2006)]{gar06}
Garc{\'\i}a P{\'e}rez, A. E., Asplund, M., Primas, F., Nissen, P. E., \&
Gustafsson, B. 2006, \aap, 451, 621 

\bibitem[Gonz{\'a}lez Hern{\'a}ndez et al.(2008)]{gon08} 
Gonz{\'a}lez Hern{\'a}ndez, J.~I., et al.\ 2008, \aap, 480, 233 

\bibitem[Israelian et al.(1998)]{isr98} 
Israelian, G., Garc{\'\i}a L\'opez, R. J., \& Rebolo, R. 1998, \apj,
507, 805 

\bibitem[Israelian et al.(2001)]{isr01} 
Israelian, G., Rebolo, R., Garc{\'\i}a L\'opez, R. J., Bonifacio,
P., Molaro, P., Basri, G., \& Shchukina, N. 2001, \apj, 551, 833

\bibitem[Kiselman(2001)]{kis01} 
Kiselman, D.\ 2001, New Astronomy Review, 45, 559 

\bibitem[Korn et al.(2003)]{kor03} 
Korn, A.~J., Shi, J., \& Gehren, T.\ 2003, \aap, 407, 691 

\bibitem[Kurucz(1993)]{kur93}
Kurucz, R. L. 1993, ATLAS9 Stellar Atmospheres Programs and 2\,\kms
Grid. (CD-ROM No. 13, Smithsonian Astrophysical Observatory,
Cambridge, MA, USA). 

\bibitem[Lai et al.(2007)]{lai07} 
Lai, D.~K., Johnson, J.~A., Bolte, M., \& Lucatello, S.\ 2007, 
\apj, 667, 1185 

\bibitem[\protect\citeauthoryear{Ludwig}{1992}]{Ludwig92} 
Ludwig, H.-G., 1992, PhDT, University of Kiel 

\bibitem[\protect\citeauthoryear{Ludwig, Jordan, \& Steffen}{1994}]{Ludwig+al94} 
Ludwig, H.-G., Jordan, S., Steffen, M.\ 1994, A\&A, 284, 105


\bibitem[Ludwig et al.(2008)]{lud08} 
Ludwig, H.-G., Gonz{\'a}lez Hern{\'a}ndez, J.~I., Behara, N., 
Caffau, E., \& Steffen, M.\ 2008, First Stars III, 990, 268 

\bibitem[Ludwig et al.(2009)]{lud09} 
Ludwig, H.-G., Caffau, E., Steffen, M., Freytag, B., Bonifacio, P., 
\& Ku{\v c}inskas, A.\ 2009, Memorie della Societa Astronomica Italiana, 
80, 711


\bibitem[Mashonkina et al.(2010)]{mas10} 
Mashonkina, L., Gehren, T., Shi, J., Korn, A., \& Grupp, F.\ 2010, 
IAU Symposium, 265, 197 

\bibitem[Nordlund(1982)]{Nordlund82} 
Nordlund, A.\ 1982, \aap, 107, 1 

\bibitem[Nissen et al.(2002)]{nis02} 
Nissen, P. E., Primas, F., Asplund, M., \& Lambert, D. L. 2002, \aap,
390, 235


\bibitem[Sbordone et al.(2004)]{sbo04} 
Sbordone, L.,  Bonifacio, P., Castelli, F., \& Kurucz, R.~L.\ 2004,
Memorie della Societa Astronomica Italiana Supplement, 5, 93  

\bibitem[Shchukina et al.(2005)]{shc05} 
Shchukina, N.~G., Trujillo Bueno, J., \& Asplund, M.\ 2005, \apj, 618,
939  

\bibitem[Sneden(1973)]{sne73}
Sneden, C. 1973,  PhD Dissertation, Univ. of Texas, Austin

\bibitem[\protect\citeauthoryear{Steffen, Ludwig, \& Freytag}{1995}]{Steffen+al95} 
Steffen, M., Ludwig, H.-G., Freytag, B.\ 1995, A\&A, 300, 473 

\bibitem[Stein \& Nordlund(1998)]{ste98} 
Stein, R.~F., \& Nordlund, A.\ 1998, \apj, 499, 914 


\bibitem[Th{\'e}venin \& Idiart(1999)]{the99} 
Th{\'e}venin, F., \& Idiart, T.~P.\ 1999, \apj, 521, 753 

\bibitem[\protect\citeauthoryear{V{\"o}gler, Bruls, \& Sch{\"u}ssler}{2004}]{Voegler+al04} 
V{\"o}gler, A., Bruls, J.~H.~M.~J., Sch{\"u}ssler, M.\ 2004, A\&A, 421,
741   

\bibitem[Wedemeyer et al.(2004)]{wed04}
Wedemeyer, S., Freytag, B., Steffen, M., Ludwig, H.-G., \& Holweger,
H.\ 2004, \aap, 414, 1121  

\end{thebibliography}
\end{document}